\documentclass[nofootinbib,twocolumn]{revtex4-1}
\usepackage[T1]{fontenc}
\usepackage{graphicx,subcaption}
\usepackage{rotating}
\usepackage{bm}        
\usepackage{amssymb}   
\usepackage{bm}
\usepackage{float}
\usepackage{tikz}
\usepackage{diagbox}
\usepackage{braket}
\usepackage{color} 
\usepackage{colortbl}
\usepackage{amsmath}
\usepackage{xcolor}
\usepackage{soul}
\usepackage{float,subcaption}
\usepackage[rightcaption]{sidecap}
\DeclareCaptionJustification{justified}{\justifying}
\definecolor{google-red}{rgb}{0.0, 0.0, 0.0}
\definecolor{red}{rgb}{0.0, 0.0, 0.0}
\definecolor{google-white}{HTML}{FFFFFF}
\definecolor{google-yellow}{HTML}{FFA700}
\definecolor{google-green}{HTML}{008744}
\definecolor{google-blue}{HTML}{0057E7}

\newcommand{\req}[1]{\begin{align}#1\end{align}}

\makeatletter

    \def\CT@@do@color{%
      \global\let\CT@do@color\relax
            \@tempdima\wd\z@
            \advance\@tempdima\@tempdimb
            \advance\@tempdima\@tempdimc
    \advance\@tempdimb\tabcolsep
    \advance\@tempdimc\tabcolsep
    \advance\@tempdima2\tabcolsep
            \kern-\@tempdimb
            \leaders\vrule
                    \hskip\@tempdima\@plus  1fill
            \kern-\@tempdimc
            \hskip-\wd\z@ \@plus -1fill }
    \makeatother

\def\k1{k_1}
\def\k2{k_2}
\def\q1{q_1}
\def\q2{q_2}

\def\({\left (}
\def\){\right )}
\def\[{\left [}
\def\]{\right ]}

\newcommand{\beq}{\begin{equation}}
\newcommand{\eeq}{\end{equation}}

\DeclareMathAlphabet\mathbfcal{OMS}{cmsy}{b}{n}

\begin{document}


\title{Compositional optimization of quantum circuits  for quantum kernels 
of support vector machines}

\author{Elham Torabian and Roman V. Krems}
 \affiliation{
Department of Chemistry, University of British Columbia, Vancouver, B.C. V6T 1Z1, Canada \\
Stewart Blusson Quantum Matter Institute, Vancouver, B.C. V6T 1Z4, Canada }



\date{\today}

\begin{abstract}
While quantum machine learning (ML) has been proposed to be one of the most promising applications of quantum computing, 
how to build quantum ML models that outperform classical ML remains a major open question. 
Here, we demonstrate a Bayesian algorithm for constructing quantum kernels for support vector machines that adapts quantum gate sequences to data. 
The algorithm increases the complexity of quantum circuits incrementally by appending quantum gates selected with
Bayesian information criterion as circuit selection metric and Bayesian optimization 
of the parameters of the locally optimal quantum circuits identified. 
The performance of the resulting quantum models for the classification problems {\color{google-red} considered here}  significantly exceeds that of optimized classical models with conventional kernels. 
\end{abstract}

\maketitle


\section{Introduction}

Quantum machine learning (QML) has recently emerged as a new research field aiming to take advantage of quantum computing for machine learning (ML) tasks. 
It has been shown that embedding data into gate-based quantum circuits can be used to produce kernels for ML models by quantum measurements \cite{schuld2021supervised,schuld2019quantum,rebentrost2014quantum, havlivcek2019supervised,mengoni2019kernel,bartkiewicz2020experimental,park2020theory, suzuki2020analysis, park2020practical,chatterjee2016generalized,glick2021covariant,blank2020quantum,wu2021application,haug2021large,otten2020quantum}.  
A major question that remains open is whether quantum kernels thus produced can outperform classical kernels for practical ML applications. 
The advantage of quantum kernels may manifest itself in reducing the computational hardness of big data problems. 
For example, it has been shown that a classification problem based on the discrete logarithm problem, which is believed to be in the NP and bounded-error quantum polynomial (BQP) complexity classes, 
can be efficiently solved with quantum kernels \cite{liu2021rigorous}. { It has also been shown that ML models with quantum kernels can efficiently solve a classification problem with the complexity of the $k$-Forrelation problem \cite{jonas}, which is proven to be promiseBQP-complete \cite{aaronson, aaronson2}. This proves that all classification problems with BQP (and promiseBQP) complexity can be solved efficiently with QML.} However, it is not yet clear if classical big data problems of practical importance can benefit from QML. 

The opposite limit is relevant for applications where observations are exceedingly expensive. For such applications, learning is based on small data. 
If quantum kernels can produce models with better inference accuracy for small data problems than classical kernels, QML may offer a practical application of quantum computing for classification, interpolation, and optimization of functions that are exceedingly expensive to evaluate. 
However, the generalization performance of quantum kernels is sensitive to the quantum circuit ansatz \cite{schuld2021supervised,cerezo2021variational,nakaji2021quantum,peters2021machine}.
The choice of the ansatz is an open problem in QML. 
For variational quantum optimization and quantum simulation applications, a common strategy is to choose a fixed ansatz and optimize the parameters of the quantum circuit \cite{schuld2021effect,mitarai2018quantum,watabe2019quantum,qi2021qtn,terashi2021event,chen2020hybrid,blance2021quantum,kwak2021introduction,sierra2020dementia,chen2020variational,lloyd2020quantum,hubregtsen2021training,henry2021quantum}. 
However, for most optimal ML models, it is necessary to develop algorithms that adapt the ansatz to data. 
Previous work has addressed this problem by optimizing gate sequences using
 quantum combs \cite{chiribella2008quantum}  for cloning, storage retrieving, discrimination, estimation, and tomographic characterization of quantum circuits,
 graph theory \cite{shafaei2013optimization} to develop quantum Fourier transforms and reversible benchmarks, quantum triple annealing minimization \cite{gyongyosi2019quantum} for quantum annealing applications, reinforcement learning \cite{fosel2021quantum} to solve the Max-Cut problem, a variable structure approach \cite{bilkis2021semi} for the variational quantum eigensolver for condensed matter and quantum chemistry applications, {an adaptive variational algorithm \cite{grimsley2019adaptive} for molecular simulations, quantum local search \cite{tomesh2021quantum} for finding the maximum independent set (MIS) of a graph}, {\color{red} layerwise learning \cite{skolik2021layerwise} for quantum neural networks}, and multi-objective genetic algorithms \cite{altares2021automatic} for quantum classification models.
 
 {The most general approach to identifying the best quantum circuit (QC) for a given ML task involves (i) search in the space of all possible permutations of quantum gates; (ii) optimization of quantum gate parameters for each QC during the search. 
 Both (i) and (ii) become quickly intractable as the number of gates increases. Both (i) and (ii) also require a circuit selection metric to guide the search and optimization. 
   {\color{google-red}
 Optimization of the architecture and parameters of quantum kernels is further complicated by the complexity of quantum measurements.   
    Estimating the kernel matrix of a ML model by measurements on a quantum computer scales with the number of training points $N$ as ${\cal O}(N^3)$ \cite{kernel-estimation-complexity}. Therefore, it is important to develop algorithms for optimization of quantum circuits with small $N$. Once found, and if robust, the quantum kernels can then be used for non-parametric models with large $N$, in order to make accurate predictions.}
  Within the Bayesian approach, ML models can be discriminated by marginal likelihood (see Section III below). 
 However, computation of marginal likelihood also becomes quickly intractable as the model complexity increases.  
 Marginal likelihood can be approximated by the Bayesian information criterion (BIC) \cite{hastie2009elements}. However, this approximation is valid in the limit of large {\color{google-red} $N$}. It is not known if BIC can be used as a QC selection metric for small data problems.

 In the present work, we demonstrate an algorithm that builds performant quantum circuits for small data classification problems using BIC as a circuit selection metric and a compositional search }that involves two stages: (i) incremental construction of the quantum circuit {by appending individual quantum gates to $K$ optimal QCs of lower complexity}; (ii)  Bayesian optimization of the parameters of {$M$ locally optimal} quantum circuits. 
{This algorithm reduces to greedy search in the limit of $K=1$ and full search of the QC space in the large $K$ and $M$ limit. The choice of $K$ thus balances the efficiency of greedy search with the volume of explored circuit space. By limiting optimization of QC parameters to $M$ locally optimal circuits, this algorithm reduces the computational complexity of simultaneous optimization of QC architecture and parameters.

}

We demonstrate the approach by building an optimal ansatz for quantum support vector machines (SVM). 
 SVM is a kernel method  that can be used for both classification and regression problems \cite{hastie2009elements}. SVM kernels allow classification of data that are not linearly separable. 
SVM is well suited for small data problems. 
The quantum analog of SVM uses quantum kernels. 
We show that the performance of quantum SVM can be systematically enhanced by building quantum kernels in a data-adaptable approach. 
In order to achieve this, we convert the output of the SVM classifier to probabilistic predictions and compute the Bayesian information criterion \cite{neath2012bayesian,shahriari2015taking}. 
Using two unrelated classification problems, we show that quantum kernels thus obtained outperform standard classical kernels and achieve enhanced inference accuracy for small data problems.  
The method proposed here is inspired by the work of Duvenaud {\it et al.} \cite{extrapolation-2}, that demonstrated the possibility of enhancing the performance of classical 
Gaussian process models through compositional kernel search guided by BIC.



\section{Quantum circuit descriptors} 
 
Supervised learning for classification uses $N$ input - output pairs $\{\bm x, y\}$, with inputs $\bm x$ represented by multi-dimensional vectors and outputs $y$ encoding the class labels. We consider binary classification problems with $y =  \{ 0, 1 \}$. 
For a QML algorithm, the input vectors $\bm x$ must be encoded into quantum states. Here, we use the following encoding of $n$-dimensional vectors $\bm x  \in \mathbb{R}^n$ into states of $n$ qubits, each initially in state $|0\rangle$:
\begin{align}
    \ket{\Phi ({\bm x})}= \exp{\left( i\displaystyle \sum_{k} x_k Z_k \right)} H^{\bigotimes n} \ket{0}^{\bigotimes n}
    \label{map}
\end{align}
where $x_k$ is the $k$-th component of vector $\bm x$, $Z_k$ is the Pauli $Z$-gate acting on $k$-th qubit, and $H$ represents the Hadamard gate putting qubits into coherent superpositions. 
The state $ \ket{\Phi ({\bm x})}$ is then operated on by a sequence of one- and two-qubit gates to produce a general quantum state ${\cal U}(\bm x)  \ket{\Phi ({\bm x})}$. The measurable quantities
\begin{align}
k(\bm x, \bm x') =    | \bra{\Phi ({\bm x'})}  {\cal U}^\dagger(\bm x') {\cal U}(\bm x)  \ket{\Phi ({\bm x})} |^2
\label{q-kernels}
\end{align}
have the properties of the kernel of a reproducing kernel Hilbert space and can thus be used as kernels of SVM.

\begin{figure}
    \centering
    \includegraphics[width=\linewidth]{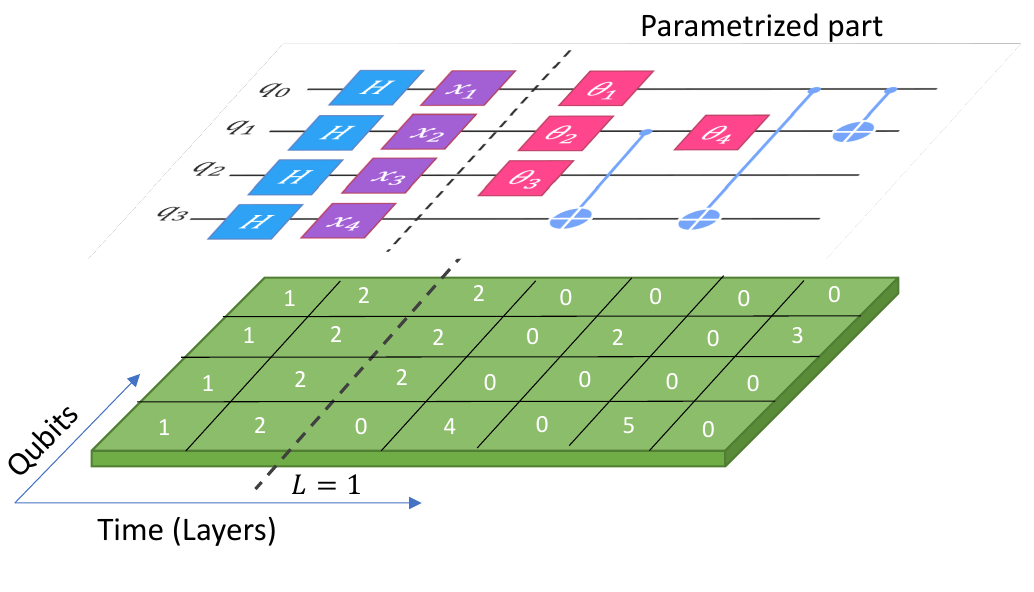}
    \caption{Schematic diagram of a quantum circuit (upper) used for quantum kernels of SVM and a matrix descriptor (lower) of the gate sequence. Each gate is assigned a numerical descriptor, encoding the type of gate and  the index of qubits entangled for CNOT gates shown by blue crosses in the upper diagram. 
    }
    \label{fig2.1}
\end{figure}

Here, we propose and demonstrate an algorithm that systematically increases the complexity of $\cal U$ to enhance the performance of quantum kernels (\ref{q-kernels}). 
As depicted in Fig. \ref{fig2.1}, we choose $\cal U$ to consist of $L$ layers, each consisting of one-qubit $R_Z$ gates and/or two-qubit CNOT gates. Each of the $R_Z$ gates for qubit $k$ is parametrized by $\theta$ as follows: 
\req{
R_{Z,k}(\theta)
=
  \begin{pmatrix}
 e^{-i \theta x_k} &
   0 \\
   0 &
   e^{i \theta x_k} 
   \end{pmatrix}.
 }
This yields a quantum circuit with the number of free parameters $\theta$ equal to the number of $R_Z$ gates. 

{
Although not essential, it is convenient for the present work to represent QCs by numerical descriptors. The numerical representation of QCs can also be used for the analysis of QCs, such as their classification into performant and non-performant kernels. 
We adopt the following approach to represent QC by numerical vectors. 
As illustrated in Fig. \ref{fig2.1} (lower part), each QC is described by a numerical matrix that encodes:  `no gate', Hadamard, CNOT, or $R_Z$. Each row and column correspond to qubits and layers, respectively. We assign $\{1, 2, 3, 4, 5\}$ to \{Hadamard, $R_Z$, CNOT$(q,q-1)$, CNOT$(q,q-2)$ and CNOT$(q,q-3)$\}, respectively. The arguments of CNOT$(a,b)$ indicate the target ($a$) and control ($b$) qubits.
The numerical descriptors 
corresponding to $n$ qubits and $g$ types of gates thus generated form a total of $[{(g+n-1)!}/{n! (g-1)!}]\times K \times L$ quantum circuits.}


\section{Marginal likelihood for QC selection}

{Within the Bayesian approach, we can assign probability $P({\cal Q}_i | {\rm Data})$ to a quantum model ${\cal Q}_i$, which can be expressed using Bayes' theorem as follows:
\req{
P({\cal Q}_i | {\rm Data}) = \frac{P({\rm Data} | {\cal Q}_i )P({\cal Q}_i)}{P({\rm Data})},
}
where ${\rm Data} = \{ {\bm X}, \bm y\}$, $\bm X$ is the data matrix and $\bm y$ is a vector of outputs in the training data set. 
For two QCs ${\cal Q}_1$ and ${\cal Q}_2$, we can write: 
\req{
\frac{P({\cal Q}_1 | {\rm Data})}{P({\cal Q}_2 | {\rm Data})} = \frac{  P({\rm Data} | {\cal Q}_1 )}{  P({\rm Data} | {\cal Q}_2 )} \times \frac{P({\cal Q}_1)}{P({\cal Q}_2)}.
}
Assuming no prior knowledge of circuits performance, we choose all $P({\cal Q}_i)$ to be the same, which shows that QCs can be discriminated by 
$P({\rm Data} | {\cal Q}_i)$. If a QC ${\cal Q}_i$ is parametrized by a set of parameters denoted collectively by $\bm \theta$, 
we can define the likelihood function for the quantum model as
\req{
{\cal L}(\bm \theta | {\rm Data}, {\cal Q}_i) = p({\rm Data} | \bm \theta, {\cal Q}_i), 
}
where $p({\rm Data} | \bm \theta, {\cal Q}_i)$ is the probability that the quantum model $ {\cal Q}_i$ with parameters $\bm \theta$ yields the observations in the training set $\{ {\bm X}, \bm y\}$. This shows that marginal likelihood
\req{
P({\rm Data} | {\cal Q}_i) = \int_{\bm \theta} p({\rm Data} | {  \bm \theta}, {\cal Q}_i) p({  \bm \theta} | {\cal Q}_i) d\bm \theta
\label{ml}
}
can be used to discriminate between quantum models with different circuit architectures. As the number of circuit parameters grows, the integrals in Eq. (\ref{ml}) become intractable. However, for a large number of training points $N$, they can be approximated as \cite{hastie2009elements}:
\req{
\log (P({\rm Data} | {\cal Q}_i)) \approx \log \hat {\cal L}({\cal Q}_i)  - \frac{d}{2}\log(N),
}
where $\hat {\cal L}({\cal Q}_i)  = {\cal L}(\bm{\hat \theta} | {\rm Data}, {\cal Q}_i)$ is the maximum value of likelihood and $d$ is the number of parameters in $\bm \theta$. 
The Bayesian information criterion (BIC) is defined as 
\begin{align}
    {\rm BIC}=- 2 \log \hat {\cal L} + d \log N,   
   \label{bic}
\end{align}
which shows that it can be used to estimate the logarithm of marginal likelihood, and hence serve as a model selection metric. In the present work, we only parametrize the $R_Z$ gates, so $d$ is equal to the total number of $R_Z$ gates in the QC. While Eq. (8) is valid for $N \rightarrow \infty$, in the present work we provide empirical evidence that BIC is a valid selection metric for quantum models for classification problems with $N \approx 100$.}

{In order to compute BIC, we convert the output $\hat f(\bm x_i)$ of SVM with quantum kernels to a posterior class probability $p_i(y = 1| \bm x_i)$, approximated by a sigmoid function:
\begin{equation}
    p_i= \frac{1}{1+\exp (a\hat{f}(\bm x_i) +b)}
    \label{Platt}
\end{equation}
The parameters $a$ and $b$ are determined by maximizing the log likelihood function $ \log \hat {\cal L}$ \cite{wu2004probability}.} For the binary classification problem with $y = \{ 0, 1 \}$,  we compute $ \log \hat {\cal L}$ as
\begin{equation}
   \log \hat {\cal L} = -  \sum_{i=1} \left [y_i\log (p_i) +(1-y_i) \log (1-p_i) \right ],
\end{equation}
where $p_i$ is the probability of observing class with label $y_i=1$. In the current implementation, SVM is trained by maximizing Lagrange dual with a training set \cite{hastie2009elements}, while $\log \hat {\cal L}$  used in Eq. (9)  is computed over a separate validation set. 
{The coefficients $a$ and $b$ are found by 4-fold cross-validation with the training set. These coefficients are then used to calculate the probabilities $p_i$ for the validation set, which yields $\log \hat {\cal L}$ for the computation of BIC.}

\section{Circuit growth algorithm}

{The purpose of the present algorithm for building QCs for quantum kernels is to approximate the search of an optimal QC architecture and an optimal set of parameters in the entire space of parametrized circuits. The pseudo-code for this algorithm is presented in the supplemental material (SM) \cite{SM}.
As seen in the pseudo-code, the algorithm is controlled by two hyperparameters: $K$ and $M$.   The limit of $M=0$ and $K \rightarrow \infty$ corresponds to the purely compositional search. When $M=K$, the algorithm optimizes the architecture and parameters of QCs simultaneously. 

The architecture of the circuit is built incrementally, starting with all possible circuits with only one layer $L=1$ of the parametrized part of QC in Fig. 1. Using BIC, we identify $K$ preferred quantum kernels with $L=1$ (i.e.  $K$ kernels yielding quantum models with the lowest BIC values). The parameters of $M \leq K$ of those QCs are optimized using Bayesian optimization (BO). Details of BO, including the choice of kernels for Gaussian process regression and parameters of the acquisition function, are given in the SM \cite{SM}. 
A new layer is then added by appending all possible combinations of gates for layer two to $K$ selected circuits with $L=1$. The algorithm then selects $K$ preferred kernels with $L=2$, and the process is iterated. This search strategy directs the algorithm to increase the complexity of quantum circuits while improving the performance of quantum kernels.}


\section{Results}
To illustrate the performance of the algorithm proposed here, we use two independent four (4D) and three (3D) dimensional classification problems. For the first example, we use the data from Ref. \cite{Jain2013} to classify halide perovskites, with the chemical formula $A_2 B B^{\prime} X_6$, into metals or non-metals. The four components of $\bm x$ correspond to the ionic radii of the four atoms $A, B, B' $ and $X$. 
The second example is a 3D synthetic data set implemented in the Qiskit ML module \cite{Qiskitadhoc}. Hereafter, we will refer to the two data sets as the perovskite dataset and the {\it ad hoc} dataset, respectively. We split each data set into $100$ random training samples, a validation set with  $100$ random samples and a test set with $1442$ and $4100$ samples for the perovskite and {\it ad hoc} datasets, respectively. The training set is used to train the SVM model with a given quantum kernel, 
and the validation set is used to select quantum circuits, while the test set is used to illustrate the final model performance. 

\subsection{Quantum circuit search convergence}

\begin{figure}[h]
    \centering
    \includegraphics[width=\linewidth]{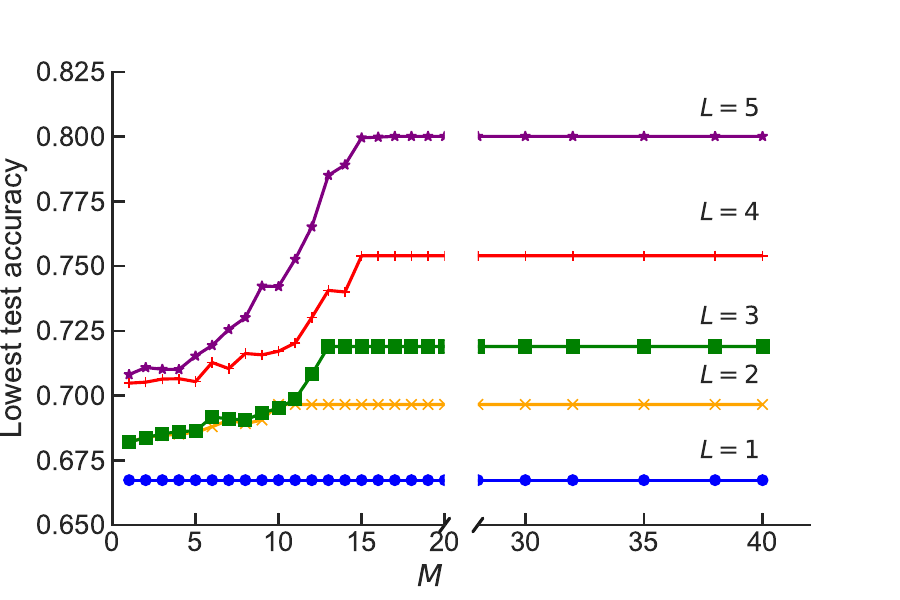}
    \caption{Convergence of Bayesian search of the quantum circuit space {\color{google-red} quantified by the lowest test accuracy (lower of TPR and TNR as defined in the text)} with respect to the algorithm hyperparameters $K$ and $M$. The results for $M \le 20$ are obtained with $K=20$. For $M\ge20$, $K=M$. }
    \label{fig:LvsM}
\end{figure}

{
Figure \ref{fig:LvsM} demonstrates the convergence of the classification accuracy with $K$ and $M$.
The convergence is examined using the lower value of the two class prediction accuracies over a test set for 
the perovskite classification problem. 
The results for $M \le 20$ are obtained with $K=20$. For $M\ge20$, $K=M$. 
The results indicate that convergence is achieved for $K = 20$ and $M=15$. 
Figure 2 shows that circuits with more layers benefit from a larger value of $M$.  
Unless stated otherwise, we use $K = M=20$ for benchmark calculations in this work. 

The results of Fig. 2 suggest interesting extensions of the present algorithm. For example, the computation complexity can be reduced by 
introducing the dependence of both $M$ and $K$ on the number of layers $L$, i.e. $M \rightarrow M(L)$ and $K \rightarrow K(L)$.
{\color{google-red} For example, the results of Fig. 2 suggest that convergence can be reached with $M=10$ for $L = 1$ and $L = 2$, $M = 13$ for $L = 3$, and $M = 15$ for $L > 3$, which requires a much smaller number of circuits to be optimized than in the present work. 
}
 To understand the relative roles of parameter vs architecture optimization, we will also compare below the performance of the fully optimized search ($K=M=20$) with the search, where the architecture of QC is optimized first and the parameters of a single best QC are optimized at the final stage of the kernel construction. 
 Models obtained using this algorithm are hereafter referred to as $M=0(1)$. 
Finally, we note that the results of Fig. 2 indicate that the algorithm monotonically converges to an optimal circuit with significantly better performance than a randomly chosen quantum kernel. 
The monotonic convergence indicates that an optimal search algorithm can be determined by increasing $K$ and $M \leq K$. 
}

\subsection{Quantum kernel selection metric} 
The performance of classification models such as SVM is generally quantified by test accuracy over a holdout set or equivalent metrics, such as the $F$ score, often $F1$ \cite{fawcett2006introduction}. 
It is convenient to label the two classes as positive and negative and define the true positive rate (TPR) and the true negative rate (TNR) as 
the ratio of correct predictions to the total number of test points for a given class. 
{\color{google-red}Unless otherwise specified,} the test/validation accuracy of the model is defined here as the balanced average of TPR and TNR computed {\color{google-red} as (TPR + TNR)/2} over the test/validation set. 
{\color{google-red} The test accuracy for a specific class refers to TPR or TNR over the test set. The lowest test accuracy refers to the lower of TPR and TNR over the test set. Since the model performance is limited by the class with the lower prediction accuracy, most of the results show the lowest test accuracy.}
More details about different model selection metrics used here are provided in SM \cite{SM}.

\begin{figure}[h]
    \centering
    \includegraphics[width=\linewidth]{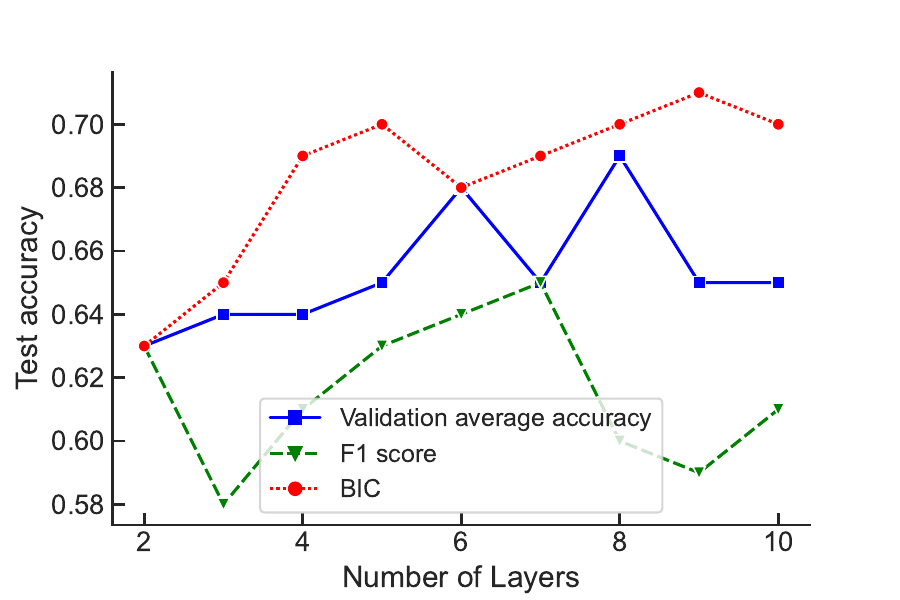} \\
     \includegraphics[width=\linewidth]{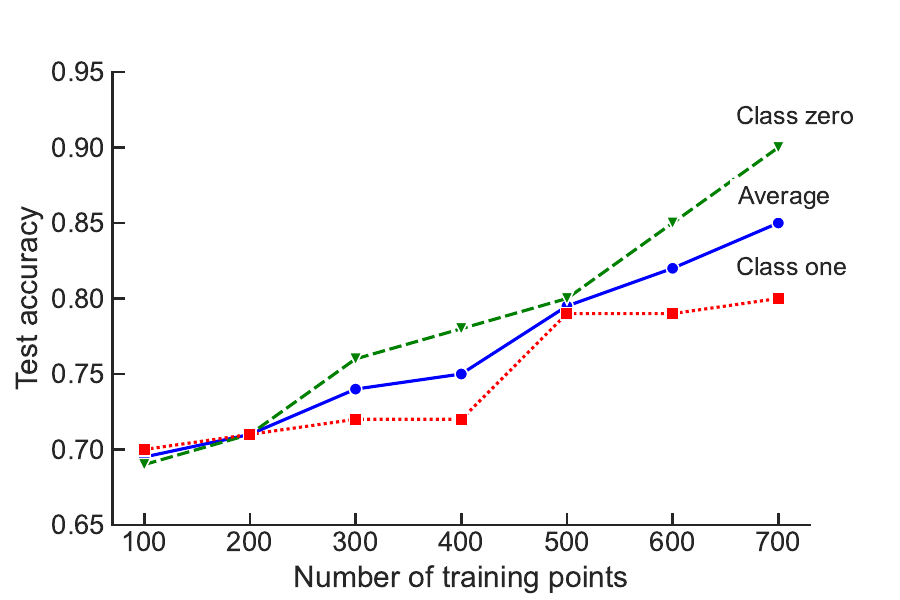}
    \caption{Upper: Quantum classification model accuracy averaged over two classes over the holdout test set as a function of the number of layers in $\cal U$ constructed using three model selection metrics: BIC (circles), validation accuracy (squares), $F1$ score (triangles). The models are trained with 100 training points. Lower: Improvement of the quantum classification accuracy with the number of training points by applying the best obtained quantum kernel using 100 training points (fixed kernel). Class-specific test accuracy represents TPR and TNR, as defined in the text. 
    }
    \label{fig2}
\end{figure}

To illustrate the importance of BIC for the present algorithm, it is instructive to compare the performance of BIC as a QC selection metric with other possible metrics. Figure \ref{fig2} uses the perovskite dataset to compare the performance of the quantum models with quantum circuits selected using three different metrics: accuracy, $F1$ score and BIC, all computed over the validation sets.  
The results demonstrate that the classification accuracy or $F1$ score cannot be used as metrics for quantum circuit selection. On the other hand, 
BIC leads to both better models for each layer $L$ and significant improvement of the quantum classification models as the kernel complexity increases. We observe a similar trend for the other data set (provided in SM \cite{SM}).

{ BIC is known to be a rigorous model selection metric asymptotically, in the limit of a large number of training points \cite{hastie2009elements}. 
For a family of $K$ models, the probability of model $m$ being selected can be estimated as:

\begin{equation}
    P(\textrm{Model}_m)=\frac{e^{{-{\rm BIC}_m}/2}}{\sum_i e^{{-{\rm BIC}_i}/2}}
\end{equation}
If the family of models includes the true model, the probability of selecting the correct model approaches one as $K \rightarrow \infty$ \cite{hastie2009elements}.
In this limit, the integral yielding marginal likelihood can be replaced with the maximized likelihood as shown in Eq. (\ref{bic}). However, Eq. (\ref{bic}) may not be valid in the small data limit. Our results in Fig. \ref{fig2} show that BIC remains an effective selection metric for quantum models trained with as few as 100 training points.

To further demonstrate the expressiveness of the quantum kernels selected by BIC, we show  in the lower panel of Fig. \ref{fig2} the improvement of the quantum classification accuracy with the number of training points for the perovskite data set. 
The models illustrated in Fig. \ref{fig2} are obtained with 100 training points using $K=20$ and $M=0$, and are subsequently used in SVM with a larger number of training points. 
The results illustrate two important points: (i) quantum kernels obtained by optimization of QC architecture with BIC based on a small number of training points produce models that generalize well for problems with larger training sets, as expected of any performant kernel -- this is a sanity test for the resulting quantum models; (ii)
performant quantum kernels can be obtained by optimization of QC architecture without optimization of QC parameters (i.e. with $M=0$).

}

\subsection{ Classical vs. quantum classification models}

\begin{figure}[h]
    \centering
    \includegraphics[width=\linewidth]{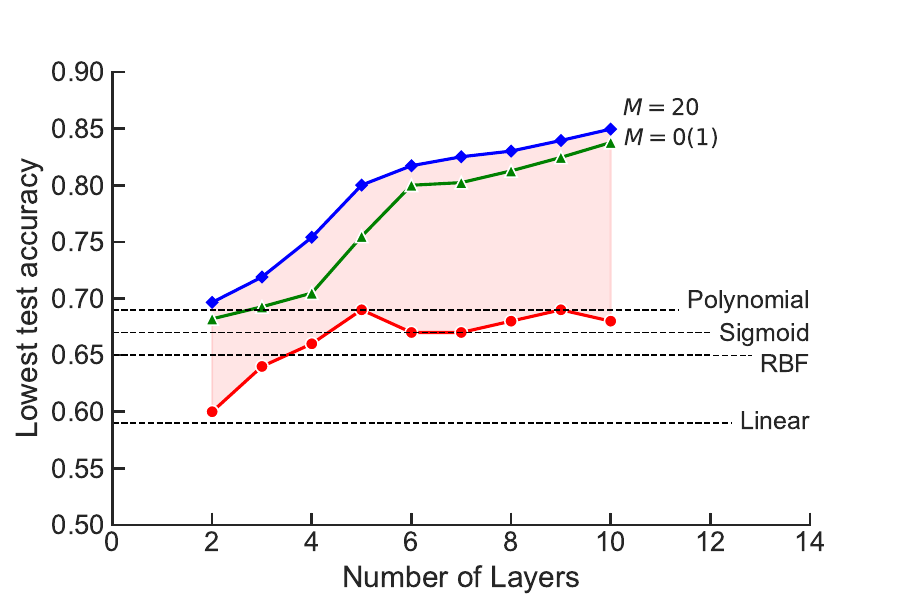} \\
     \includegraphics[width=\linewidth]{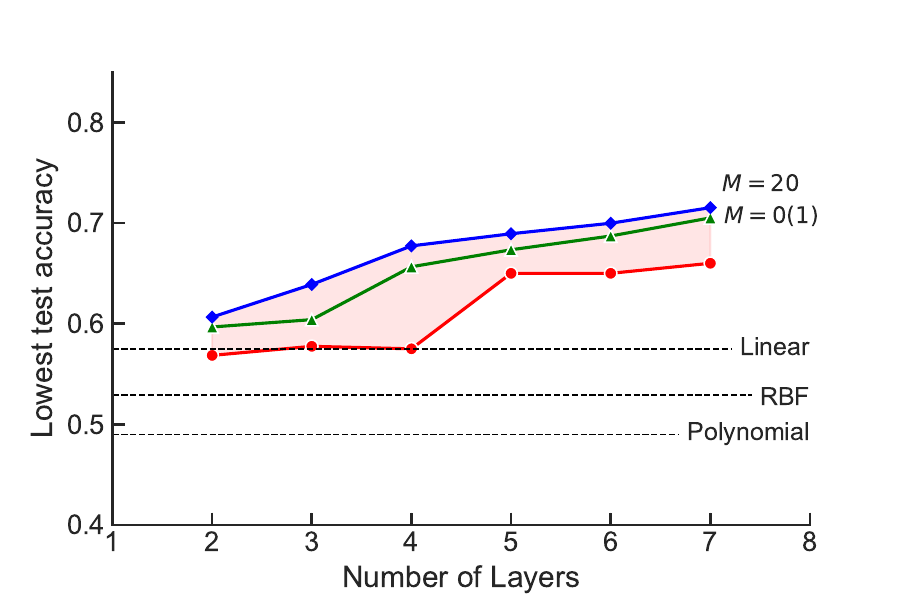}
    \caption{Lowest prediction accuracy ({\color{google-red} lower of TPR and TNR as defined in the text}) evaluated on the holdout test set for the perovskite (upper) and {\it ad hoc} (lower) classification problems. The horizontal broken lines show the best accuracy for SVM models with the classical kernels indicated as labels. 
   The classical kernels are optimized using the validation set. The symbols show the results of the best quantum model for each layer $L$: circles -- quantum kernels constructed using Bayesian search of QC architecture with fixed parameters of all quantum gates; 
 triangles -- the $K=20, M= 0 (1)$ search, where architecture is optimized without any optimization of circuit parameters, but the parameters of the best QC are optimized to enhance performance;
   diamonds -- quantum circuits from the $K = M = 20$ Bayesian search involving simultaneous optimization of both the QC architecture and QC parameters. 
       }
    \label{fig4}
\end{figure}

Figure \ref{fig4} compares the results of the predictions of the best quantum models with the classical models based on commonly used optimized kernels. 
Specifically, we train four classical models with the radial basis function (RBF), linear, cubic polynomial and sigmoid kernel functions.    
The models are trained with the same number and distribution of training points as for the quantum models and the parameters of each kernel function are optimized to minimize the error over identical validation sets for all models.

{There are three sets of quantum models shown:
\begin{itemize}
\item
 First, we construct the best quantum kernel for each number of layers $L$, without any optimization of the circuit parameters. These models correspond to Bayesian search hyperparameters $K =20$ and $M=0$ and illustrate the improvement of quantum kernels that can be achieved by optimizing only the circuit architecture. The prediction accuracy of SVM models with quantum kernels thus obtained is shown by circles. The circles in Fig. \ref{fig4} thus represent the results of purely compositional optimization. 

\item Second, we vary the parameters $\theta$ of all $R_Z$ gates in the best quantum kernel for a given $L$ to minimize error on the validation set. 
These models correspond to Bayesian search hyperparameters $K =20$ and $M=0$, but include parameter optimization for one best circuit at each layer. Hence, we label these results, shown by 
triangles in Fig. 4, as $M=0 (1)$.  
The purpose of this calculation is to demonstrate the improvement due to optimization of the parameters of the locally optimal circuits. The triangles in Fig. \ref{fig4} thus represent the results of purely compositional optimization followed by optimization of parameters of the best QC of a given complexity.

\item Finally, the diamonds in Fig. 4 show the results of the fully optimized search with $K = M = 20$. The diamonds in Fig. \ref{fig4} thus represent the results of simultaneous optimization of QC architecture and parameters.

\end{itemize}

}

For a binary classification problem, one can evaluate the model test accuracy for each class separately. 
The performance of classification models is generally limited by the lower of the two prediction accuracies.
Figure \ref{fig4} shows the lowest test accuracy of the best models trained with $N=100$.  It can be seen that the performance of quantum models significantly exceeds the prediction accuracy achievable with classical kernels considered. 
SM presents the detailed comparisons of classical and quantum model predictions for each class separately. 

{ 
It is interesting to observe that purely compositional optimization followed by optimization of parameters of the best QC (triangles in Fig. \ref{fig4}) produces quantum kernels yielding test errors comparable with the test errors of the full 
simultaneous optimization of QC architecture and parameters (diamonds in Fig. \ref{fig4}). Furthermore, the difference between the diamonds and the triangles in Fig. \ref{fig4} decreases with increasing QC complexity. 
This trend is observed for both classification problems considered here. 
This suggests that the most optimal algorithm for constructing complex quantum kernels should separate the compositional optimization and the optimization of QC parameters. 
}

\section{Conclusion}

We have demonstrated an algorithm for constructing quantum kernels for quantum SVM that adapts gate sequences and gate parameters to data for classification problems. The quantum kernels for SVM are built through compositional search in the space of gate permutations and Bayesian optimization of quantum circuits with locally optimal architecture. The search algorithm is controlled by two hyperparameters ($K$ and $M$) that balance the greediness, and hence the efficiency, of compositional search with the volume of quantum circuit space explored. The algorithm is designed to approach the complete search of the entire space of optimized quantum circuits in the asymptotic limit of large $K$ and $M$. We have shown that quantum kernels with enhanced generalization performance can be obtained through search restricted to a small fraction ($< 0.5 \%$ for the present examples) of the total space.

We have illustrated the performance of the resulting quantum kernels by comparisons with classical models based on several conventionally used kernels for two non-linear classification problems. Our results show that optimized quantum kernels yield models with significantly better classification accuracy than the classical models considered here. It is important to note that we do not claim quantum advantage for the quantum kernels obtained in this work. Although the parameters of the classical kernels were optimized, we did not optimize the functional form of the classical models. More importantly, we consider kernels based on a small number of qubits. Such kernels are classically simulable. However, the present work offers a new approach to identify QML models with enhanced performance, potentially making QML useful for practical applications, including for interpolation, classification and optimization of objective functions that are exceedingly difficult to evaluate. We also conjecture that if quantum kernels can be constructed to be expressive for low-dimensional problems, they can also be constructed to be expressive for high-dimensional problems, for which quantum kernels are not classically simulable with existing classical hardware resources.

Critical to any algorithm of automated quantum ansatz construction is the quantum circuit selection metric. Our results show that performant quantum kernels for classification problems can be selected by Bayesian information criterion, while the validation error and $F1$ score fail as quantum model selection metrics for the problems considered here. The choice of the ansatz remains an open problem in QML. The present work offers an algorithm to identify subdomains of quantum gate permutations that yield good kernels and bad kernels. This can potentially be used to identify the essential elements of quantum circuits that are responsible for the expressiveness of the ensuing kernels.

\section*{Acknowledgments}

This work was supported by NSERC of Canada.


\clearpage
\newpage

\begin{thebibliography}{99}
\bibitem{schuld2021supervised}
M. Schuld, {Supervised quantum machine learning models are kernel methods}, {\it arXiv:2101.11020} (2021).

\bibitem{schuld2019quantum}
M. Schuld and N. Killoran, {Quantum machine learning in feature hilbert spaces}, {\it Phys. Rev. Lett.} {\bf 122}, 040504 (2019).

\bibitem{rebentrost2014quantum}
P. Rebentrost, M. Mohseni, and S. Lloyd, {Quantum support vector machine for big data classification}, {\it Phys. Rev. Lett.} {\bf 113}, 130503 (2014).

\bibitem{havlivcek2019supervised}
V. Havl{\'\i}{\v{c}}ek, A. D. C{\'o}rcoles, K. Temme, A. W. Harrow, A. Kandala, J. M. Chow, and J. M. Gambetta, {Supervised learning with quantum-enhanced feature spaces}, {\it Nature} {\bf 567}, 209 (2019).

\bibitem{mengoni2019kernel}
R. Mengoni and A. Di Pierro, {Kernel methods in quantum machine learning}, {\it Quantum Mach. Intell.} {\bf 1}, 65 (2019).

\bibitem{bartkiewicz2020experimental}
K. Bartkiewicz, C. Gneiting, A. {\v{C}}ernoch, K. Jir{\'a}kov{\'a}, K. Lemr, and F. Nori, {Experimental kernel-based quantum machine learning in finite feature space}, {\it Sci. Rep.} {\bf 10}, 1 (2020).

\bibitem{park2020theory}
D. K. Park, C. Blank, and F. Petruccione, {The theory of the quantum kernel-based binary classifier}, {\it Phys. Lett. A} {\bf 384}, 126422 (2020).

\bibitem{suzuki2020analysis}
Y. Suzuki, H. Yano, Q. Gao, S. Uno, T. Tanaka, M. Akiyama, and N. Yamamoto, {Analysis and synthesis of feature map for kernel-based quantum classifier}, {\it Quantum Mach. Intell.} {\bf 2}, 1 (2020).

\bibitem{park2020practical}
J. E. Park, B. Quanz, S. Wood, H. Higgins, and R. Harishankar, {Practical application improvement to Quantum SVM: theory to practice}, {\it arXiv:2012.07725} (2020).

\bibitem{chatterjee2016generalized}
R. Chatterjee and T. Yu, {Generalized coherent states, reproducing kernels, and quantum support vector machines}, {\it arXiv:1612.03713} (2016).

\bibitem{glick2021covariant}
J. R. Glick, T. P. Gujarati, A. D. Corcoles, Y. Kim, A. Kandala, J. M. Gambetta, and K. Temme, {Covariant quantum kernels for data with group structure}, {\it arXiv:2105.03406} (2021).

\bibitem{blank2020quantum}
C. Blank, D. K. Park, J. K. K. Rhee, and F. Petruccione, {Quantum classifier with tailored quantum kernel}, {\it npj Quantum Inf.} {\bf 6}, 1 (2020).

\bibitem{wu2021application}
S. L. Wu, S. Sun, W. Guan, C. Zhou, J. Chan, C. L. Cheng, T. Pham, Y. Qian, A. Z. Wang, R. Zhang, and M. Livny, {Application of quantum machine learning using the quantum kernel algorithm on high energy physics analysis at the LHC}, {\it Phys. Rev. Res.} {\bf 3}, 033221 (2021).

\bibitem{haug2021large}
T. Haug, C. N. Self, and M. S. Kim, {Large-scale quantum machine learning}, {\it arXiv:2108.01039} (2021).

\bibitem{otten2020quantum}
M. Otten, I. R. Goumiri, B. W. Priest, G. F. Chapline, and M. D. Schneider, {Quantum machine learning using gaussian processes with performant quantum kernels}, {\it arXiv:2004.11280} (2020).

\bibitem{liu2021rigorous}
Y. Liu, S. Arunachalam, and K. Temme, {A rigorous and robust quantum speed-up in supervised machine learning}, {\it Nat. Phys.} {\bf 17}, 1013 (2021).

\bibitem{jonas}
J. J\"{a}ger and R V. Krems, {Universal expressiveness of variational quantum classifiers and quantum kernels for support vector machines}, {\it arXiv:2207.05865} (2022). 

\bibitem{aaronson}
S. Aaronson, {BQP and the polynomial hierarchy}, {\it Proc. Annu. ACM Symp. Theory Comput.}, 141 (2010).

\bibitem{aaronson2} 
S. Aaronson and A. Ambainis, {Forrelation: A problem that optimally separates quantum from classical computing}, {\it Proc. Annu. ACM Symp. Theory Comput.}, 307 (2015).

\bibitem{cerezo2021variational}
M. Cerezo, A. Arrasmith, R. Babbush, S. C. Benjamin, S. Endo, K. Fujii, J. R. McClean, K. Mitarai, X. Yuan, L. Cincio, and P. J. Coles, {Variational quantum algorithms}, {\it Nat. Rev. Phys.} {\bf 3}, 625 (2021).

\bibitem{nakaji2021quantum}
K. Nakaji, H. Tezuka, and N. Yamamoto, {Quantum-enhanced neural networks in the neural tangent kernel framework}, {\it arXiv:2109.03786} (2021).

\bibitem{peters2021machine}
E. Peters, J. Caldeira, A. Ho, S. Leichenauer, M. Mohseni, H. Neven, P. Spentzouris, D. Strain, and G. N. Perdue, {Machine learning of high dimensional data on a noisy quantum processor}, {\it npj Quantum Inf.} {\bf 7}, 1 (2021).

\bibitem{schuld2021effect}
M. Schuld, R. Sweke, and J. J. Meyer, {Effect of data encoding on the expressive power of variational quantum-machine-learning models}, {\it Phys. Rev. A} {\bf 103}, 032430 (2021).

\bibitem{mitarai2018quantum}
K. Mitarai, M. Negoro, M. Kitagawa, and K. Fujii, {Quantum circuit learning}, {\it Phys. Rev. A} {\bf 98}, 032309 (2018).

\bibitem{watabe2019quantum}
M. Watabe, K. Shiba, M. Sogabe, K. Sakamoto, and T. Sogabe, {Quantum circuit parameters learning with gradient descent using backpropagation}, {\it arXiv:1910.14266} (2019).

\bibitem{qi2021qtn}
J. Qi, C. H. H. Yang, and P. Y. Chen, {QTN-VQC: An End-to-End Learning framework for Quantum Neural Networks}, {\it arXiv:2110.03861} (2021).

\bibitem{terashi2021event}
K. Terashi, M. Kaneda, T. Kishimoto, M. Saito, R. Sawada, and J. Tanaka, {Event classification with quantum machine learning in high-energy physics}, {\it Comput. Softw. Big Sci.} {\bf 5}, 1 (2021).

\bibitem{chen2020hybrid}
S. Y. C. Chen, C. M. Huang, C. W. Hsing, and Y. J. Kao, {Hybrid quantum-classical classifier based on tensor network and variational quantum circuit}, {\it arXiv:2011.14651} (2020).

\bibitem{blance2021quantum}
A. Blance and M. Spannowsky, {Quantum machine learning for particle physics using a variational quantum classifier}, {\it J. High Energy Phys.} {\bf 2021}, 1 (2021).

\bibitem{kwak2021introduction}
Y. Kwak, W. J. Yun, S. Jung, J. K. Kim, and J. Kim, {Introduction to quantum reinforcement learning: Theory and pennylane-based implementation}, {\it Int. Conf. ICT Converg.}, 416 (2021).

\bibitem{sierra2020dementia}
D. Sierra-Sosa, J. Arcila-Moreno, B. Garcia-Zapirain, C. Castillo-Olea, and A. Elmaghraby, {Dementia prediction applying variational quantum classifier}, {\it arXiv:2007.08653} (2020).

\bibitem{chen2020variational}
S. Y. C. Chen, C. H. H. Yang, J. Qi, P. Y. Chen, X. Ma, and H. S. Goan, {Variational quantum circuits for deep reinforcement learning}, {\it IEEE Access} {\bf 8}, 141007 (2020).

\bibitem{lloyd2020quantum}
S. Lloyd, M. Schuld, A. Ijaz, J. Izaac, and N. Killoran, {Quantum embeddings for machine learning}, {\it arXiv:2001.03622} (2020).

\bibitem{hubregtsen2021training}
T. Hubregtsen, D. Wierichs, E. Gil-Fuster, P. J. H. Derks, P. K. Faehrmann, and J. J. Meyer, {Training quantum embedding kernels on near-term quantum computers}, {\it arXiv:2105.02276} (2021).

\bibitem{henry2021quantum}
L. P. Henry, S. Thabet, C. Dalyac, and L. Henriet, {Quantum evolution kernel: Machine learning on graphs with programmable arrays of qubits}, {\it Phys. Rev. A} {\bf 104}, 032416 (2021).

\bibitem{chiribella2008quantum}
G. Chiribella, G. M. D’Ariano, and P. Perinotti, {Quantum circuit architecture}, {\it Phys. Rev. Lett.} {\bf 101}, 060401 (2008).

\bibitem{shafaei2013optimization}
A. Shafaei, M. Saeedi, and M. Pedram, {Optimization of quantum circuits for interaction distance in linear nearest neighbor architectures}, {\it 2013 50th ACM/EDAC/IEEE Design Automation Conference (DAC)}, 1 (2013).

\bibitem{gyongyosi2019quantum}
L, Gyongyosi and S. Imre, {Quantum circuit design for objective function maximization in gate-model quantum computers}, {\it Quantum Inf. Process.} {\bf 18}, 1 (2019).

\bibitem{fosel2021quantum}
T. F{\"o}sel, M. Y. Niu, F. Marquardt, and L. Li, {Quantum circuit optimization with deep reinforcement learning}, {\it arXiv:2103.07585} (2021).

\bibitem{bilkis2021semi}
M. Bilkis, M. Cerezo, G. Verdon, P. J. Coles, and L. Cincio, {A semi-agnostic ansatz with variable structure for quantum machine learning}, {\it arXiv:2103.06712} (2021).

\bibitem{grimsley2019adaptive}
H. R. Grimsley, S. E. Economou, E. Barnes, and N. J. Mayhall, {An adaptive variational algorithm for exact molecular simulations on a quantum computer}, {\it Nat. Commun.} {\bf 10}, 1 (2019).

\bibitem{tomesh2021quantum}
T. Tomesh, Z. H. Saleem, and M. Suchara, {Quantum local search with quantum alternating operator ansatz}, {\it arXiv:2107.04109} (2021).
{\color{red}
\bibitem{skolik2021layerwise}
A. Skolik, J. R. McClean, M. Mohseni, P. van der Smagt, and M. Leib, {Layerwise learning for quantum neural networks}, {\it Quantum Mach. Intell.}, {\bf 3}, 1 (2021).}

\bibitem{altares2021automatic}
S. Altares-L{\'o}pez, A. Ribeiro, and J. J. Garc{\'\i}a-Ripoll, {Automatic design of quantum feature maps}, {\it Quantum Sci. Technol.} {\bf 6}, 045015 (2021).


{\color{google-red}
\bibitem{kernel-estimation-complexity}
T. Hubregtsen, D. Wierichs, E. Gil-Fuster, P. J. H. Derks, P. K. Faehrmann, and J. J. Meyer, {Training quantum embedding kernels on near-term quantum computers}, {\it Phys. Rev. A}  {\bf 106}, 042431 (2022).
}

\bibitem{hastie2009elements}
T. Hastie, R. Tibshirani, and J. H. Friedman, {The elements of statistical learning: data mining, inference, and prediction}, {\it Springer} {\bf 2} (2009).

\bibitem{neath2012bayesian}
A. A. Neath and J. E. Cavanaugh, {The Bayesian information criterion: background, derivation, and applications}, {\it Wiley Interdiscip. Rev. Comput. Stat.} {\bf 4}, 199 (2012).

\bibitem{extrapolation-2}
D. K. Duvenaud, J. Lloyd,  R. Grosse,  J. B. Tenenbaum,  and Z. Ghahramani, {Structure discovery in nonparametric regression through compositional kernel search},
{\it Proceedings of the 30th International Conference on Machine Learning Research} {\bf 28}, 1166 (2013).


\bibitem{shahriari2015taking}
B. Shahriari, K. Swersky, Z. Wang, R. P. Adams, and N. De Freitas, {Taking the human out of the loop: A review of Bayesian optimization}, {\it Proc. IEEE} {\bf 104}, 148 (2015).

\bibitem{wu2004probability}
T. F. Wu, C. J. Lin, and R. Weng, {Probability estimates for multi-class classification by pairwise coupling}, {\it Adv. Neural Inf. Process. Syst.} {\bf 16} (2003).


\bibitem{Jain2013}
A. Jain,S. P. Ong, G. Hautier, W. Chen, W. D. Richards, S. Dacek, S. Cholia, D. Gunter, D. Skinner, G. Ceder, and K. A. Persson, {The Materials Project: A materials genome approach to accelerating materials innovation}, {\it APL Mater.} {\bf 1}, 011002 (2013).

\bibitem{Qiskitadhoc}
Qiskit Machine Learning Development Team, {Qiskit machine learning}, \\ {\it https://qiskit.org/documentation/machine-learning} (2021).

\bibitem{fawcett2006introduction}
T. Fawcett, {An introduction to ROC analysis}, {\it Pattern Recognit. Lett.} {\bf 27}, 861 (2006).

\bibitem{khan2016descriptors}
A. U. Khan, {Descriptors and their selection methods in QSAR analysis: paradigm for drug design}, {\it Drug Discov. Today} {\bf 21}, 1291 (2016).

\bibitem{kamath2017application}
V. Kamath and A. Pai, {Application of Molecular Descriptors in Modern Computational Drug Design-An Overview}, {\it Res. J. Pharm. Technol.} {\bf 10}, 3237 (2017).

\bibitem{xue2000molecular}
L. Xue and J. Bajorath, {Molecular descriptors in chemoinformatics, computational combinatorial chemistry, and virtual screening}, {\it Comb. Chem. High Throughput Screen.} {\bf 3}, 363 (2000).

\bibitem{amigo2009review}
J. M. Amig{\'o}, J. G{\'a}lvez, and V. M. Villar, {A review on molecular topology: applying graph theory to drug discovery and design}, {\it Sci. Nat.} {\bf 96} (2009).

\bibitem{todeschini2008handbook}
R. Todeschini and V. Consonni, {Handbook of molecular descriptors}, {\it John Wiley \& Sons} (2008).

\bibitem{seo2020development}
M. Seo, H. K. Shin, Y. Myung, S. Hwang, and K. T. No, {Development of Natural Compound Molecular Fingerprint (NC-MFP) with the Dictionary of Natural Products (DNP) for natural product-based drug development}, {\it J. Cheminform.} {\bf 12}, 1 (2020).

\bibitem{RDKit}
G. Landrum, {RDKit: Open-Source Cheminformatics Software}, {\it http://www.rdkit.org/}.

\bibitem{SM}
Supplemental Material


\end{thebibliography}
\end{document}


\title{Supplemental Material for\\
Compositional optimization of quantum circuits  for quantum kernels 
of support vector machines}

\author{Elham Torabian and Roman V. Krems}
 \affiliation{Department of Chemistry, University of British Columbia, Vancouver, B.C. V6T 1Z1, Canada \\
Stewart Blusson Quantum Matter Institute, Vancouver, B.C. V6T 1Z4, Canada }

\date{\today}

\maketitle

 The following is the pseudocode for the algorithm of this work:

\begin{algorithm}
	\caption{Quantum kernel optimization for SVM}
	\raggedright{\textbf{Input:}\\ (i) Classification data set $\{ \bm x, y \}$: including training, validation; \\ (ii) A set of quantum gates; \\ (iii) $L_{\rm max}$: number of quantum circuit (QC) layers; \\ (iv) $K$: number of QCs for local search; \\ (v) $M$: number QCs with optimized parameters; \\ (vi) ${\cal N}$: maximum number of iterations in Bayesian optimization (BO).\\}
    \raggedright{\textbf{Output:} An optimized QC architecture with optimized gate parameters $\bm \theta^*$}
	\begin{algorithmic}[1]
		\State Build a QC that encodes inputs into gate parameters
		\State Initialize list OptimalQC with $K$ identical QCs
		\For {All $1< L \leq L_{\rm max}$}
		\For {Each element in OptimalQC}
		\For {All possible gate combinations in layer $L$}
		\State Append a layer of gates to QC 
		\State Compute  the quantum kernel  $k(\bm x, \bm x^{\prime})$
		\State Train an SVM model with $k(\bm x, \bm x^{\prime})$
		\State Convert outputs of SVM to probabilistic predictions
		\State Calculate BIC with the validation set
		\EndFor
		\EndFor
		\State Sort all resulting quantum kernels by BIC in increasing order
		\State Replace list OptimalQC with $K$ lowest BIC entries
		\For{$M \leq K$ QCs with lowest BIC $\in$ OptimalQC}
		\While{ iteration $\leq {\cal N}$}
		\State Determine the gate parameters $\bm \theta$ using the acquisition function of BO
		\EndWhile
		\State Assign the resulting gate parameters to $\bm \theta^\ast$
		\EndFor
        \EndFor
	\end{algorithmic} 
\end{algorithm} 

\section{\label{sm:2}Bayesian Optimization}

Bayesian Optimization (BO) is a gradient-free optimization approach that can be used to identify the global maximizer (or minimizer) of a non-convex black-box function (unknown objective function $f$) \cite{shahriari2015taking}. In this work, BO is applied to optimize the parameters $\bm \theta$ of a quantum circuit. 
The objective function for this optimization is BIC$=f(\bm \theta)$ computed over a validation set.
BO is initialized by evaluations of BIC at $50$ randomly selected combinations of parameters $\bm \theta$. A Gaussian process (GP) model $\mathcal{F}(\bm \theta)$ is trained using the results of these evaluations. Each GP model $\mathcal{F}$ is 
characterized by the mean  $\mu(\bm \theta) $ and the uncertainty $\sigma(\bm \theta)$, which are functions of $\bm \theta$.  The subsequent evaluations of BIC are performed at the maximum of the acquisition function $\alpha(\bm \theta)$ defined as

\begin{equation}
    \alpha(\bm \theta)=\mu(\bm \theta)+\kappa \sigma(\bm \theta) \; ,
    \label{eq.1}
\end{equation}
\noindent
where $\kappa$ is a hyperparameter that determines the balance between exploration and exploitation. At each iteration, we add the result of the new evaluation to the previous sample of $\bm \theta$ and re-train the GP model. The value of $\kappa$ in Eq. (\ref{eq.1}) is chosen to be $\kappa=1$.
We use the RBF kernel for the GP models $\mathcal{F}$ and run BO for $\sim 10 \times d$ iterations sampling $ d \in {1,...,16}$ dimensions of the $\bm \theta$ parameter space to identify the optimal parameters $\bm \theta^*$. 
{\color{red}
\section{\label{sm:3} Datasets}

The present algorithm is demonstrated using two different classification problems with four and three dimensions of the input space. The first problem categorizes halide perovskites with the chemical formula $A_2BB^{\prime}X_6$ as metals or non-metals based on the ionic radii of the atoms $A$, $B$, $B^{\prime}$, and $X$. The second dataset is a synthetic 3D dataset ass provided by the Qiskit ML module \cite{Qiskitadhoc}. These datasets are referred to as the perovskite dataset and the {\it ad hoc} dataset, respectively.

\begin{figure}[H]
    \savebox{\imagebox}{\includegraphics[width=0.48\textwidth]{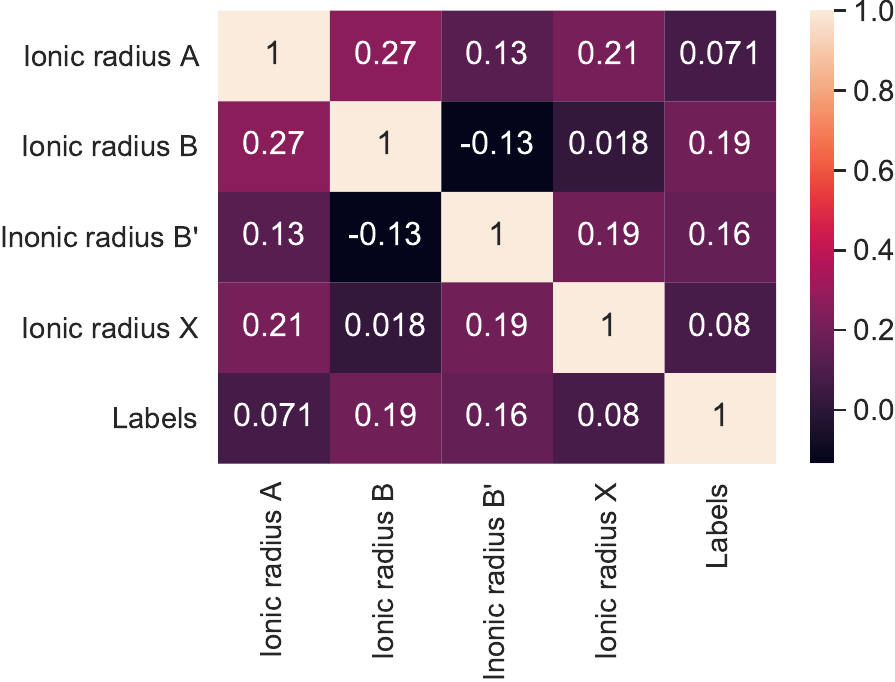}}
    \begin{subfigure}{\textwidth}
        \centering\usebox{\imagebox}
    \end{subfigure}\qquad
    \begin{subfigure}{\textwidth}
        \centering\raisebox{\dimexpr.2\ht\imagebox}{ \includegraphics[width=0.48\textwidth]{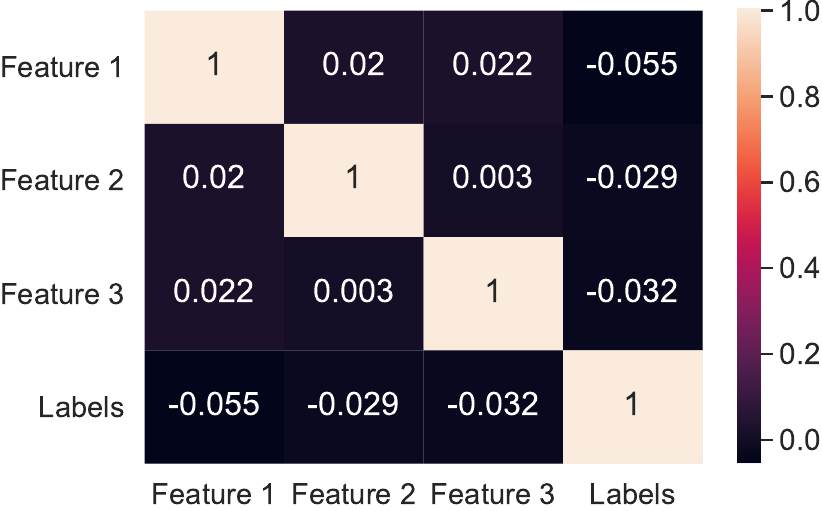}}
    \end{subfigure}
    \caption{Correlation matrices for the perovskite dataset (left) and the {\it ad hoc} dataset (right.)}
    \label{fig:corr}
\end{figure}

 Figure \ref{fig:corr} presents the correlation matrices for these datasets, illustrating the absence of significant pairwise correlations between the input variables. 
Figure \ref{fig:dist} further displays the individual and pairwise distribution plots for the perovskite and {\it ad hoc} datasets. Figure \ref{fig:dist} indicates the absence of clear pairwise correlations between the input variables. This suggests that these problems are non-trivial classification problems and should be challenging to classify with limited data, even with non-linear kernels. This is also supported by the results shown in Figure 3 of the main text, which demonstrate poor performance of the models with conventional classical kernels.}

\begin{figure}[H]
    \centering
    \includegraphics[width=0.6\textwidth]{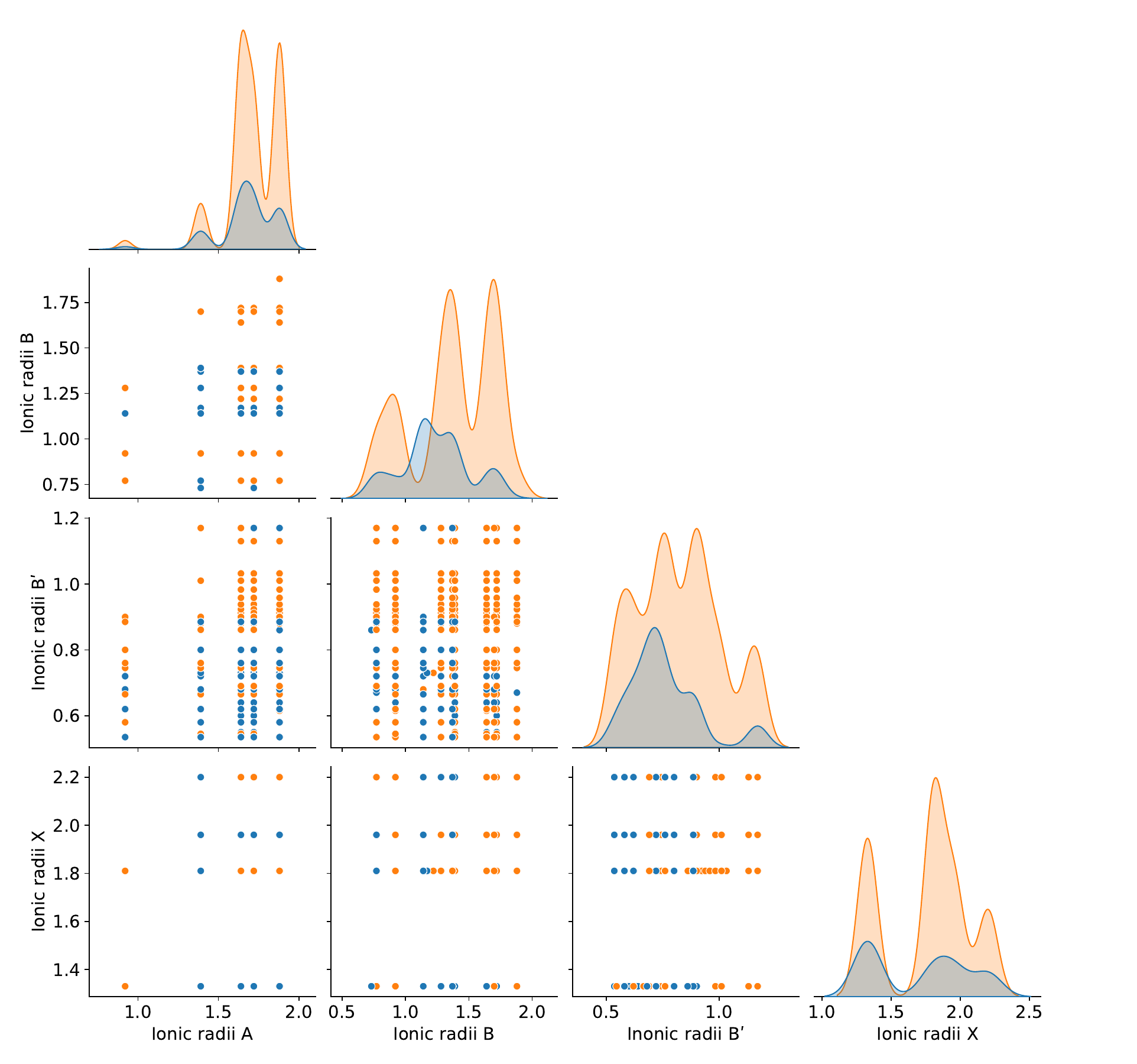}
    \includegraphics[width=0.6\textwidth]{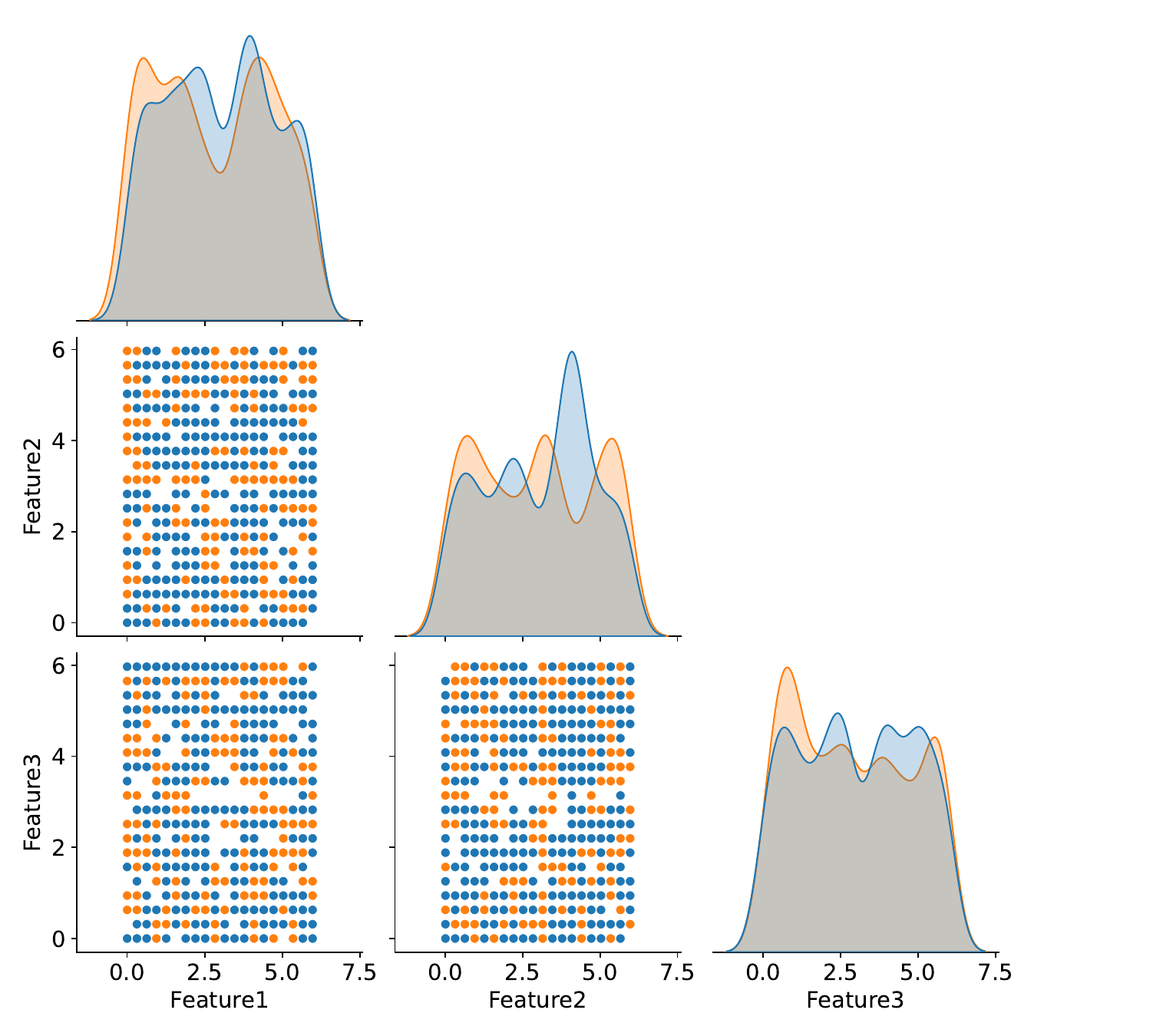}
    \caption{Individual and pairwise distribution of features for the perovskite dataset (upper panels) and the {\it ad hoc} dataset (lower panels). The blue and orange points and curves correspond to class zero samples and class one samples, respectively.}
    \label{fig:dist}
\end{figure}

\section{\label{sm:4} Model selection criteria}

For binary classification, one can characterize model accuracy by true positive rate (TPR), true negative rate (TNR), and by the average accuracy defined in the present work as (TPR + TNR)/2. TPR is defined as the ratio of correct predictions to the total number of test points for class zero, and TNR is defined in the same way for class one, which can be calculated as

\begin{equation}
    \textrm{TPR}=\frac{\textrm{TP}}{\textrm{P}}=\frac{\textrm{TP}}{\textrm{TP}+\textrm{FN}} \;,
    \label{eq:TPR}
\end{equation}

\begin{equation}
    \textrm{TNR}=\frac{\textrm{TN}}{\textrm{N}}=\frac{\textrm{TP}}{\textrm{TN}+\textrm{FP}} \;,
    \label{eq:TNR}
\end{equation}
where P, N, TP, TN, FN, and FP refer to total samples with the positive label, total samples with the negative label, true predicted samples with the positive label, true predicted samples with the negative labels, false predicted samples with the negative label, and false predicted samples with the positive label, respectively. We refer to TPR as test accuracy for class one and TNR as test accuracy for class zero. The main text presents the results for the class for which the models are least accurate (lowest average accuracy).

Another commonly used measure of accuracy for classification problems is the $F$ score, specifically $F1$, which can be defined as

\begin{equation}
    F1=\frac{\textrm{TP}}{\textrm{TP}+\frac{1}{2}(\textrm{FP}+\textrm{FN})}\; .
    \label{eq:F1score}
\end{equation}

\begin{figure}[H]
    \centering
    \includegraphics[width=0.48\linewidth]{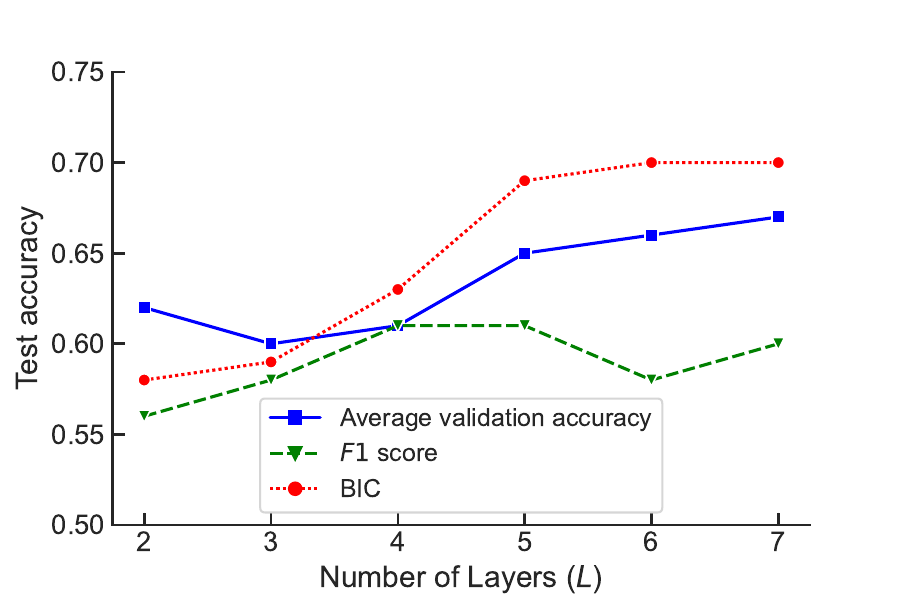}
    \includegraphics[width=0.48\linewidth]{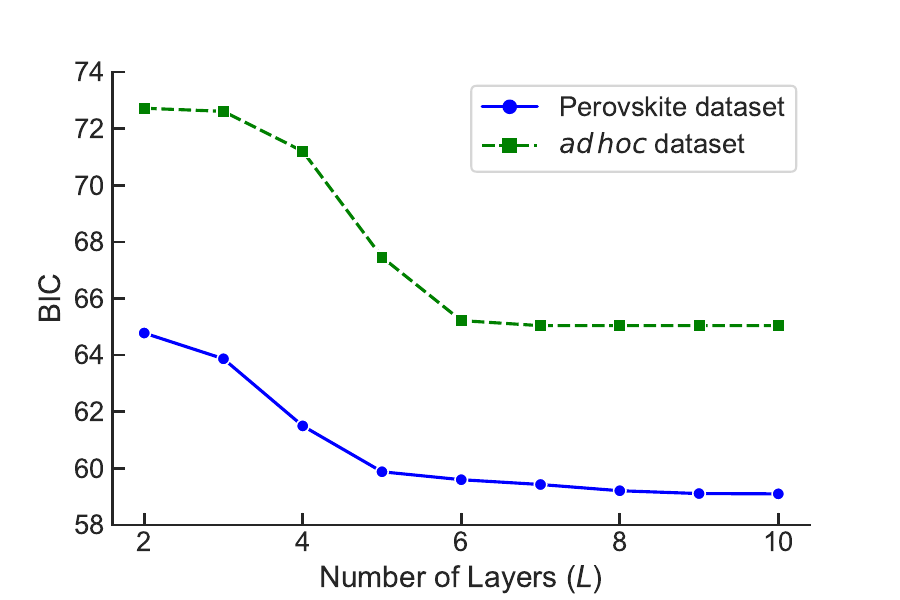} 
    \caption{Left: Quantum classification model accuracy averaged over two classes over the holdout test set as a function of the number of layers in $\cal U$ constructed using three model selection metrics: BIC (circles), validation accuracy (squares), $F1$ score (triangles). The models are trained with 100 training points. Right: Lowest value of  BIC for the classification model trained with 100 training points vs the complexity of QC ($L$).}
    \label{fig3}
\end{figure}

Figure \ref{fig3} (left panel) uses the \textit{ad hoc} dataset to compare the performance of the quantum models with quantum circuits selected using three different metrics: validation accuracy, $F1$ score, and BIC, all computed over the validation sets. The results are consistent with those in Fig. 3 (upper panel) of the main text for the perovskite data set.
Figure \ref{fig3} (right panel) shows that BIC decreases monotonically with the increasing number of QC layers, exhibiting a more significant drop at $L=1-6$ and converging to a minimum at $L=8$.

\section{\label{sm:5}Class-specific test accuracy}
 
\begin{figure}[H]
  \centering
    \includegraphics[width=0.48\linewidth]{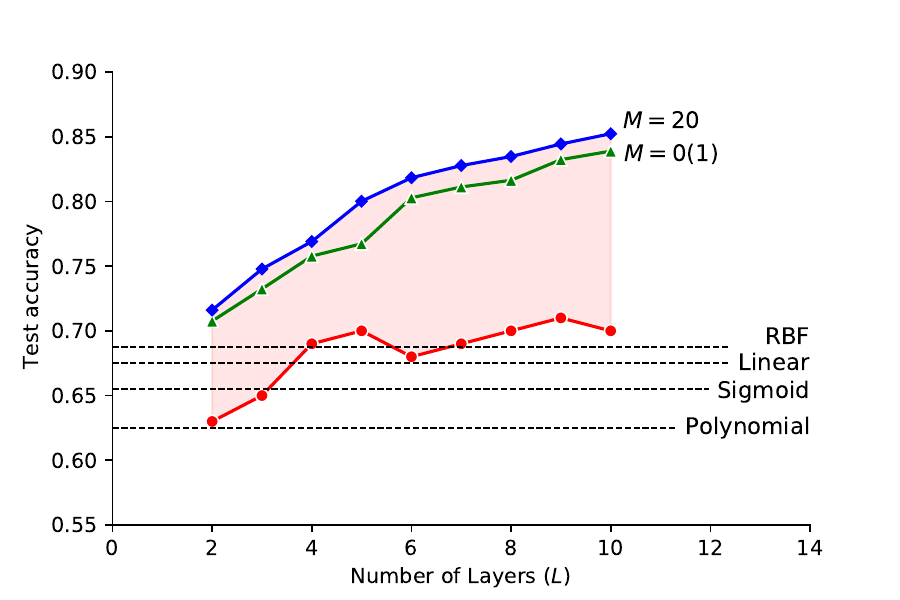}
    \includegraphics[width=0.48\linewidth]{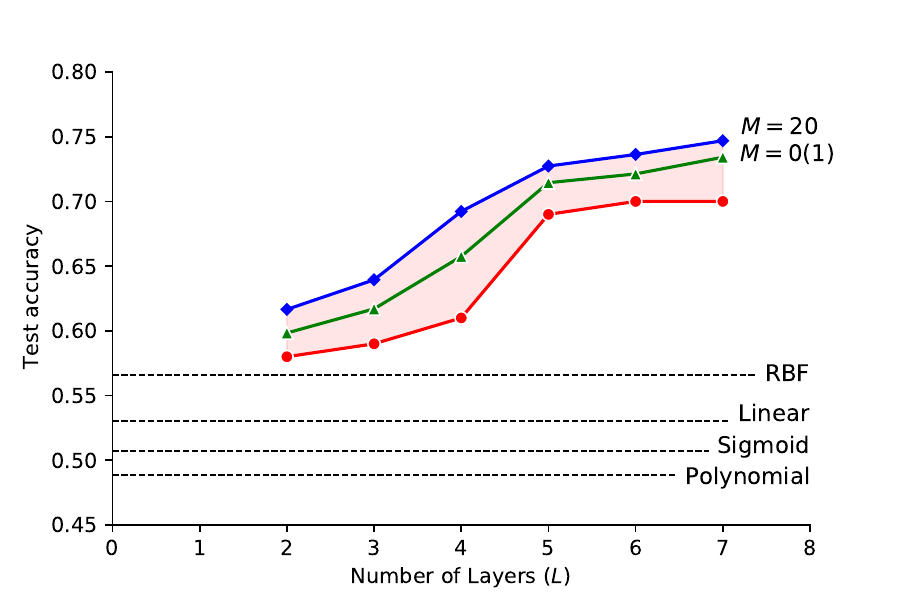}
    \caption{Test accuracy evaluated on the holdout test set for the perovskite (left) and {\it ad hoc} (right) classification problems. The horizontal broken lines show the best accuracy for SVM models with the classical kernels indicated as labels. The classical kernels are optimized using the validation set. The symbols show the results of the best quantum model for each layer $L$: circles -- quantum kernels constructed using compositional search as described in the main text with fixed parameters of all quantum gates; triangles -- the $K=20, M= 0 (1)$ search, where architecture is optimized without any optimization of circuit parameters, but the parameters of the best QC are optimized to enhance performance; diamonds -- quantum circuits from the $K = M = 20$ Bayesian search involving simultaneous optimization of both the QC architecture and QC parameters.}
    \label{fig:sm.1}
\end{figure}

\begin{figure}[H]
  \centering
    \includegraphics[width=0.48\linewidth]{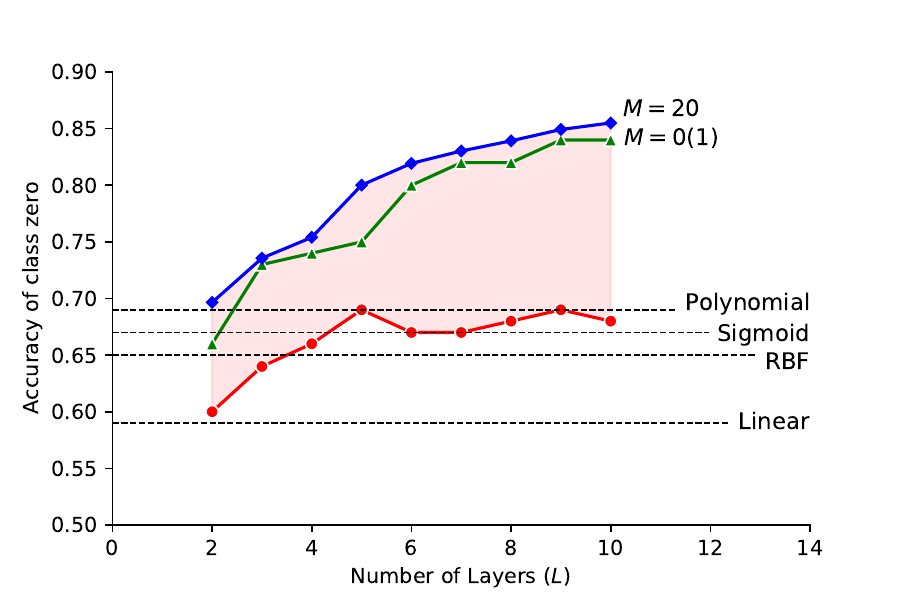}
    \includegraphics[width=0.48\linewidth]{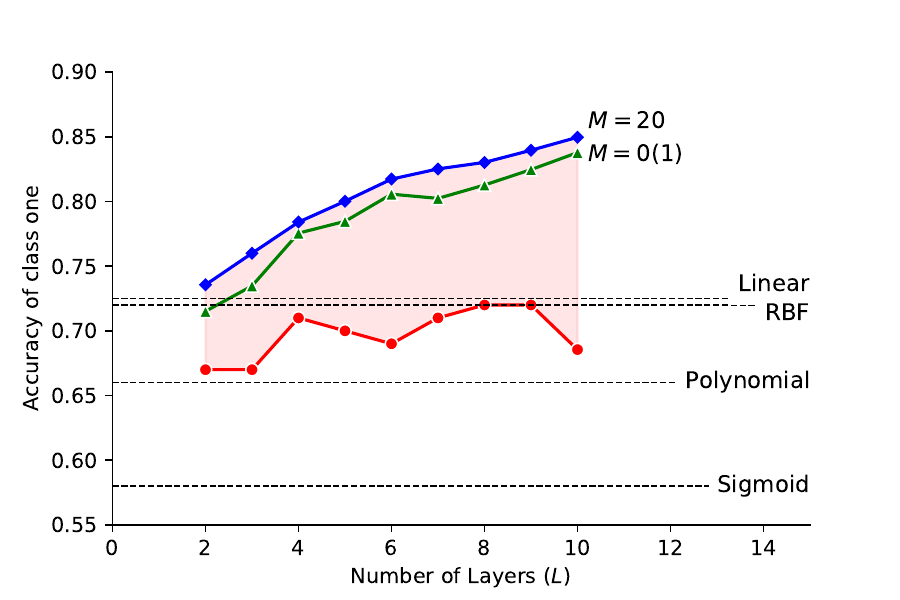}
    \caption{Test accuracy of the perovskite classification problem for the class zero (left) and class one  (right) sets. The horizontal dashed lines show the best accuracy for SVM models with the classical kernels indicated as labels that they are optimized using the validation set. The symbols show the results of the best quantum model for each layer $L$: circles -- quantum kernels constructed using compositional search as described in the main text with fixed parameters of all quantum gates; triangles -- the $K=20, M= 0 (1)$ search, where architecture is optimized without any optimization of circuit parameters, but the parameters of the best QC are optimized to enhance performance; diamonds -- quantum circuits from the $K = M = 20$ Bayesian search involving simultaneous optimization of both the QC architecture and QC parameters.}
    \label{fig:sm.2}
\end{figure}

\begin{figure}[H]
  \centering
    \includegraphics[width=0.48\linewidth]{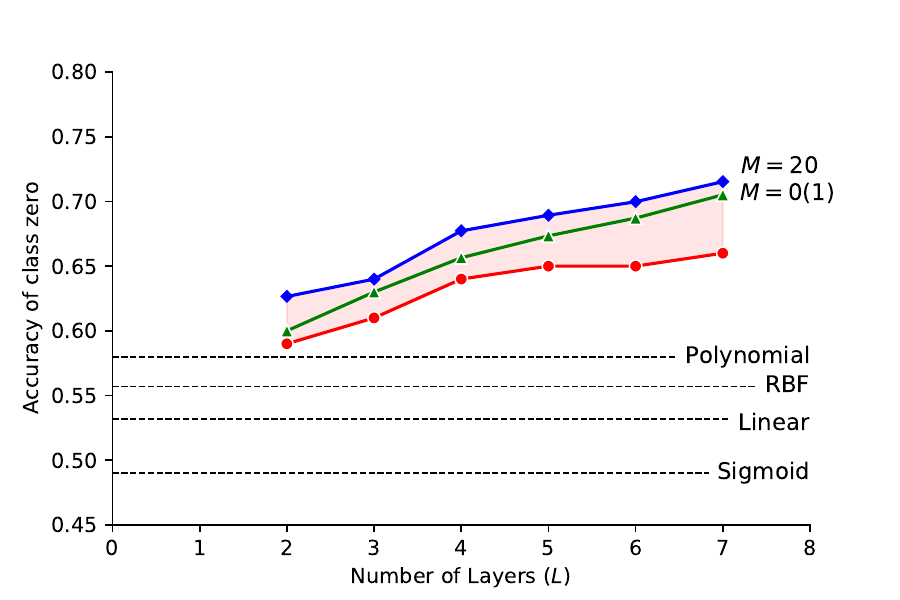}
    \includegraphics[width=0.48\linewidth]{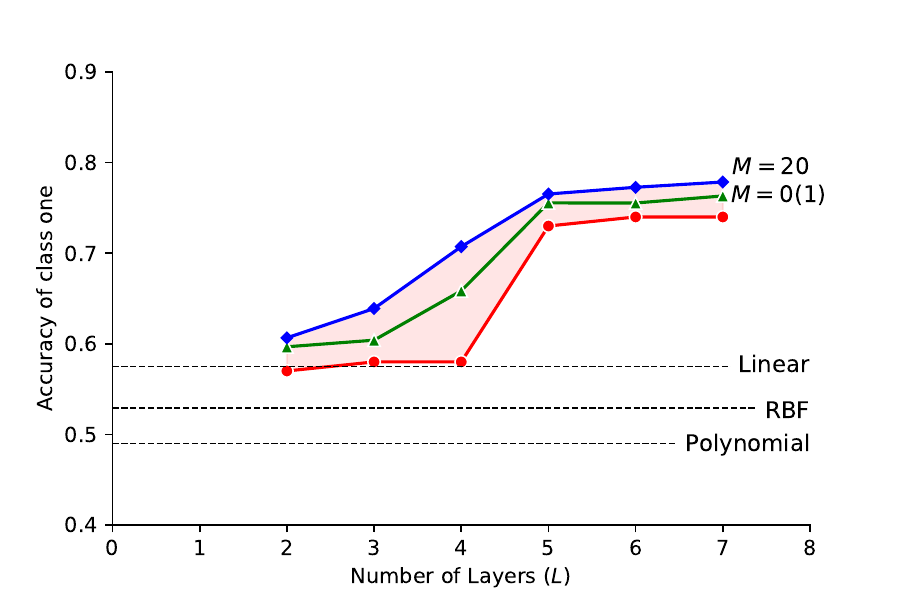}
    \caption{Test accuracy of the {\it ad hoc} classification problem for the class zero (left) and class one  (right) sets. The horizontal dashed lines show the best accuracy for SVM models with the classical kernels indicated as labels that they are optimized using the validation set. The symbols show the results of the best quantum model for each layer $L$: circles -- quantum kernels constructed using compositional search as described in the main text with fixed parameters of all quantum gates; triangles -- the $K=20, M= 0 (1)$ search, where architecture is optimized without any optimization of circuit parameters, but the parameters of the best QC are optimized to enhance performance; diamonds -- quantum circuits from the $K = M = 20$ Bayesian search involving simultaneous optimization of both the QC architecture and QC parameters.}
    \label{fig:sm.3}
\end{figure}

For completeness, we present here the average test accuracy (Figure \ref{fig:sm.1}) and the test accuracy for each class separately (Figures \ref{fig:sm.2} and  \ref{fig:sm.3}) for the two classification problems considered in this work.

